# Dark energy as the source of the time dependend Einstein cosmological constant

Miroslaw Kozlowski

Physics Department, Warsaw University , Warsaw ,Poland

Janina Marciak-Kozlowska

Institute of Electron Technology, Warsaw, Poland

Abstract

In this paper we calculate the Einstein cosmological constant.Following our results presented in monograph: From quark to bulk matter, we obtain the new formula for the time dependent cosmological constant, $\Lambda$. Following the general formula for time dependent $\Lambda$, we can describe the history of the cosmological constant

# Introduction

In this paper following the results obtained in our monograph:From quark to bulk matter , we develope the model for the time dependent Einstein cosmological constant $\Lambda$. We argue that the dark energy can be approximated as the result the dark force with the coupling constant –G, G is the Newton`s gravitational constant. As the result we obtain the new formula for the $\Lambda$

$$\Lambda = 3/(4\pi N + 3\pi) L_p c t \qquad (I1)$$

In formula (I1 ) Lp is the Planck length, cis the vacuum ligt velocity, t-time and N is the natural numer. The calculated values : R-Universe radius, v- velocity of the Universe expansion and t-time are presented in Table1

TABLE 1, MAIN RESULTS

| T[$\tau_{Planck}$] | R[m] | v[c] | $\Lambda$[m^-2] |
|---|---|---|---|
| 5 | $10^{-35}$ | 0.8 | $10^{68}$ |
| $10^{20}$ | $10^{-15}$ | 0.8 | $<10^{33}$ |
| $10^{60}$ | $10^{25}$ | 0.8 | $<10^{-47}$ |

# 1
## Universal Relaxation Processes

The differential equations of thermal energy transfer should be hyperbolic so as to exclude action at distance; yet the equations of irreversible thermodynamics – those of Navier – Stokes and Fourier are parabolic.

In the description of the evolution of any physical system, it is mandatory to evaluate, as accurately as possible, the order of magnitude of different characteristic time scales, since their relationship with the time scale of observation (the time during which we assume our description of the system to be valid) will determine, along with the relevant equations, the evolution pattern. Take a forced damped harmonic oscillator and consider its motion on a time scale much larger than both the damping time and the period of the forced oscillation. Then, what one observes is just a harmonic motion. Had we observed the system on a time scale of the order of (or smaller) than the damping time, the transient regime would have become apparent. This is rather general and of a very relevant interest when dealing with dissipative systems. It is our purpose here, by means of examples and arguments related to a wide class of phenomena, to emphasize the convenience of resorting to hyperbolic theories when dissipative processes, either outside the steady-state regime or when the observation time is of the order or shorter than some characteristic time of the system, are under consideration. Furthermore, as it will be mentioned below, transient phenomena may affect the way in which the system leaves the equilibrium, thereby affecting the future of the system even for time scales much larger than the relaxation time.

Parabolic theories of dissipative phenomena have long and a venerable history and proved very useful especially in the steady-state regime [5]. They exhibit however some undesirable features, such as acausality (see e.g., [5]), that prompted the formulation of hyperbolic theories of dissipation to get rid of them. This was achieved at the price of extending the set of field variables by including the dissipative fluxes (heat current, non-equilibrium stresses and so on) at the same footing as the classical ones (energy densities, equilibrium pressures, etc), thereby giving rise to a set of more physically satisfactory (as they much better conform with experiments) but involved theories from the mathematical point of view. These theories have the additional advantage of being backed by statistical fluctuation theory, kinetic theory of gases (Grad's 13-moment approximation), information theory and correlated random walks

A key quantity in these theories is the relaxation time $\tau$ of the corresponding dissipative process. This positive-definite quantity has a distinct physical meaning, namely the time taken by the system to return spontaneously to the steady state (whether of thermodynamic equilibrium or not) after it has been suddenly removed from it. It is, however, connected to the mean collision time tc of the particles responsible for the dissipative process It is therefore appropriate to interpret the relaxation time as the time taken by the corresponding dissipative flow to relax to its steady value.

In the book [5] the new hyperbolic non – Fourier equation for heat transport was formulated and solved.

The excitation of matter on the nuclear and atomic level leads to transfer of energy. The response of the chunk of matter (nucleus, atom) is governed by the relaxation time.

In this paper we develop the general, universal definition of the relaxation time, which depends on coupling constans .It occurs that the general formula for the relaxation time can be written as

$$\tau_i = \frac{\hbar}{m_i (\alpha_i c)^2}$$

(1)

where $m_i$ is the heat carrier mass, $\alpha_i = \left(i = e, 1/137\right)$ is coupling constant for electromagnetic interaction, c is the vacuum light speed. As the c is the maximal velocity all relaxation time fulfils the inequality

$$\tau > \tau_i$$

Consequently $\tau_i$ is the minimal universal relaxation time.

After the standards of time and space were defined the laws of classical physics relating such parameters as distance, time, velocity, temperature are assumed to be independent of accuracy with which these parameters can be measured. It should be noted that this assumption does not enter explicitly into the formulation of classical physics. It implies that together with the assumption of existence of an object and really independently of any measurements (in classical physics) it was tacitly assumed that there was a possibility of an unlimited increase in accuracy of measurements. Bearing in mind the "atomicity" of time i.e. considering the smallest time period, the Planck time, the above statement is obviously not true. Attosecond electron pulses are at the limit of time resolution. In this paragraph, we develop and solve the quantum relativistic heat transport equation for attosecond electron transport phenomena where external forces exist [5]. In paragraph 2 we develop the new hyperbolic heat transport equation which generalizes the Fourier heat transport equation for the rapid thermal processes. The hyperbolic heat transport equation (HHT) for the fermionic system has be written in the form (3):

$$\frac{1}{\left(\frac{1}{3}v_F^2\right)}\frac{\partial^2 T}{\partial t^2} + \frac{1}{\tau\left(\frac{1}{3}v_F^2\right)}\frac{\partial T}{\partial t} = \nabla^2 T$$

(3)

where T denotes the temperature, $\tau$ the relaxation time for the thermal disturbance of the fermionic system, and $v_F$ is the Fermi velocity.

In what follows we present the formulation of the HHT, considering the details of the two fermionic systems: electron gas in metals .

For the electron gas in metals, the Fermi energy has the form

$$E_F^e = (3\pi)^2 \frac{n^{2/3}\hbar^2}{2m_e},$$

(4)

where n denotes the density and me electron mass. Considering that

$$n^{-1/3} \sim a_B \sim \frac{\hbar^2}{me^2},$$

(5)

and aB = Bohr radius, one obtains

$$E_F^e \sim \frac{n^{2/3}\hbar^2}{2m_e} \sim \frac{\hbar^2}{ma^2} \sim \alpha^2 m_e c^2,$$

(6)

where c = light velocity and $\alpha$ = 1/137 is the fine-structure constant for electromagnetic interaction. For the Fermi momentum $p_F$ we have

$$p_F^e \sim \frac{\hbar}{a_B} \sim \alpha m_e c,$$

(7)

and, for Fermi velocity $\upsilon_F$,

$$\upsilon_F^e \sim \frac{p_F}{m_e} \sim \alpha c.$$

(8)

Formula (8) gives the theoretical background for the result presented in paragraph 2. Considering formula (8), equation HHT can be written as

$$\frac{1}{c^2}\frac{\partial^2 T}{\partial t^2} + \frac{1}{c^2\tau}\frac{\partial T}{\partial t} = \frac{\alpha^2}{3}\nabla^2 T.$$

(9)

As seen from (9), the HHT equation is a relativistic equation, since it takes into account the finite velocity of light.

In the following, the procedure for the quantization of temperature $T(\vec{r},t)$ in hot fermion gas will be developed. First of all, we introduce the reduced de Broglie wavelength

$$\lambda_B^e = \frac{\hbar}{m_e \upsilon_h^e}, \quad \upsilon_h^e = \frac{1}{\sqrt{3}}\alpha c,$$

(10)

and the mean free path $\lambda^e$

$$\lambda^e = \upsilon_h^e \tau^e,$$

(11)

In view of formulas (10) and (11), we obtain the HHC for electron and nucleon gases

$$\frac{\lambda_B^e}{\upsilon_h^e}\frac{\partial^2 T}{\partial t^2} + \frac{\lambda_B^e}{\lambda^e}\frac{\partial T}{\partial t} = \frac{\hbar}{m_e}\nabla^2 T^e,$$

(12)

Equations (11) and (12) are the hyperbolic partial differential equations which are the master equations for heat propagation in Fermi electron and nucleon gases. In the following, we will study the quantum limit of heat transport in the fermionic systems. We define the quantum heat transport limit as follows:

$$\lambda^e = \lambdabar_B^e,$$

(13)

In that case, Eq. (11) has the form

$$\tau^e \frac{\partial^2 T^e}{\partial t^2} + \frac{\partial T^e}{\partial t} = \frac{\hbar}{m_e}\nabla^2 T^e,$$

(14)

where

$$\tau^e = \frac{\hbar}{m_e (v_h^e)^2},$$  (15)

Equations (14) and (15) define the master equation for quantum heat transport (QHT). Having the relaxation times $\tau^e$ and one can define the "pulsations" $\omega_h^e$

$$\omega_h^e = (\tau^e)^{-1},$$  (16)

or

$$\omega_h^e = \frac{m_e (v_h^e)^2}{\hbar},$$

i.e.,

$$\omega_h^e \hbar = m_e (v_h^e)^2 = \frac{m_e \alpha^2}{3} c^2,$$

(17)

The formulas (17) define the Planck-Einstein relation for heat quanta $E_h^e$ and $E_h^N$

$$E_h^e = \omega_h^e \hbar = m_e (v_h^e)^2,$$

(18)

The heat quantum with energy $E_h = \hbar \omega$ can be named the heaton, in complete analogy to the phonon, magnon, roton, etc. For $\tau^e \to 0$, Eq. (14) are the Fourier equations with quantum diffusion coefficients $D^e$ and

$$\frac{\partial T^e}{\partial t} = D^e \nabla^2 T^e, \qquad D^e = \frac{\hbar}{m_e},$$

(19)

For finite $\tau^e$, for $\Delta t < \tau^e$, , Eq. (14) can be written as

$$\frac{1}{(v_h^e)^2}\frac{\partial^2 T^e}{\partial t^2}=\nabla^2 T^e,$$

(20)

Equations (19) and (20) are the wave equations for quantum heat transport (QHT)

It is quite interesting that the *Proca* type equation can be obtained for thermal phenomena. In the following starting with the hyperbolic heat diffusion equation the Proca equation for thermal processes will be developed [5].

When the external force is present F(x,t) the forced damped heat transport is obtained [5] (in one dimensional case):

$$\frac{1}{v^2}\frac{\partial^2 T}{\partial t^2}+\frac{m_0\gamma}{\hbar}\frac{\partial T}{\partial t}+\frac{2Vm_0\gamma}{\hbar^2}T-\frac{\partial^2 T}{\partial x^2}=F(x,t).$$

(22)

The hyperbolic relativistic quantum heat transport equation, (22), describes the forced motion of heat carriers which undergo scattering ($\frac{m_0\gamma}{\hbar}\frac{\partial T}{\partial t}$ term) and are influenced by the potential term ($\frac{2Vm_o\gamma}{\hbar^2}T$).

Equation (22) is the Proca thermal equation and can be written as [5]:

$$\left(\bar{\Box}^2+\frac{2Vm_0\gamma}{\hbar^2}\right)T+\frac{m_0\gamma}{\hbar}\frac{\partial T}{\partial t}=F(x,t),$$

$$\bar{\Box}^2=\frac{1}{v^2}\frac{\partial^2}{\partial t^2}-\frac{\partial^2}{\partial x^2}.$$

(24)

We seek the solution of equation (24) in the form

$$T(x,t)=e^{-t/2\tau}u(x,t),$$

(25)

where $\tau_i=\frac{\hbar}{mv^2}$ is the relaxation time. After substituting equation (25) in equation (24) we obtain a new equation

$$\left(\bar{\Box}^2+q\right)u(x,t)=e^{t/2\tau}F(x,t)$$

(26)

and

$$q = \frac{2Vm}{\hbar^2} - \left(\frac{m\upsilon}{2\hbar}\right)^2,$$

(27)

$$m = m_0 \gamma.$$

(28)

In free space i.e. when F(x,t) → 0 equation (24) reduces to

$$\left(\overline{\Box}^2 + q\right) u(x,t) = 0,$$

(42)

which is essentially the free *Proca* type equation.

The Proca equation describes the interaction of the attosecond electron pulse with the matter. As was shown in book [5] the quantization of the temperature field leads to the heatons – quanta of thermal energy with a mass $m_h = \frac{\hbar}{\tau \upsilon_h^2}$ [5], where $\tau$ is the relaxation time and $\upsilon_h$ is the finite velocity for heat propagation. For $\upsilon_h \to \infty$, i.e. for $c \to \infty, m_0 \to 0$, it can be concluded that in non-relativistic approximation (c = infinite) the Proca equation is the diffusion equation for massless photons and heatons.

___________________________________

Time dependend cosmological constant Λ

In the recent years the growing interest for the source of Universe expansion is observed. After the work of Supernova detecting groups the consensus for the acceleration of the moving of the space time is established [1,2]. One of the promised candidate for the repulsive is the dark energy. In this paragraph we approximate the dark energy with dark energy force=gravity with negative G

We will study the influence of the repulsive gravity ($G < 0$) on the temperature field in the universe and cosmological constant Λ. To that aim we will apply the quantum hyperbolic heat transfer equation (QHT) formulated in our earlier papers [3,4].

When substitution $G \rightarrow -G$ is performed in QHT the Schrödinger type equation is obtained for the temperature field. In papers [3,4] the quantum heat transport equation in a Planck Era was formulated:

$$\tau \frac{\partial^2 T}{\partial t^2} + \frac{\partial T}{\partial t} = \frac{(\hbar/2\pi)}{M_P} \nabla^2 T. \qquad (1)$$

In equation (1) $\tau = ((\hbar/2\pi) G/c^5)^{1/2}$ is the relaxation time, $M_P = ((\hbar/2\pi) c/G)^{1/2}$ is the mass of the Planck particle, $(\hbar/2\pi)$, $c$ are the Planck constant and light velocity respectively and *G* is the gravitational constant. The crucial role played by gravity (represented by *G* in formula (1)) in a Planck Era was investigated in paper [4]. For a long time the question whether, or not the fundamental constant of nature *G* vary with time has been a question of considerable interest. Since P. A. M. Dirac [5] suggested that the gravitational force may be weakening with the expansion of the Universe, a variable *G* is expected in theories such as the Brans-Dicke scalar-tensor theory and its extension [6,7]. Recently the problem of the varying *G* received renewed attention in the context of extended inflation cosmology [8]. It is now known, that the spin of a field (electromagnetic, gravity) is related to the nature of the force: fields with odd-integer spins can produce both attractive and repulsive forces; those with even-integer spins such as scalar and tensor fields produce a purely attractive force. Maxwell's electrodynamics, for instance can be described as a spin one field. The force from this field is attractive between oppositely charged particles and repulsive between similarly charged particles.

The integer spin particles in gravity theory are like the graviton, mediators of forces and would generate the new effects. Both the graviscalar and the graviphoton are expected to have the rest mass and so their range will be finite rather than infinite. Moreover, the graviscalar will produce only attraction, whereas the graviphoton effect will depend on whether the interacting particles are alike or different. Between matter and matter (or antimatter and antimatter) the graviphoton will produce repulsion. The existence of repulsive gravity forces can to some extent explains the early expansion of the Universe [5].

In this paper we will describe the influence of the repulsion gravity on the quantum thermal processes in the universe. To that aim we put in equation (1) $G \rightarrow -G$. In that case the new equation is obtained, viz.

$$i(h/2\pi) \frac{\partial T}{\partial t} = \left(\frac{(h/2\pi)^3 |G|}{c^5}\right)^{1/2} \frac{\partial^2 T}{\partial t^2} - \left(\frac{(h/2\pi)^3 |G|}{c}\right)^{1/2} \nabla^2 T. \qquad (2)$$

For the investigation of the structure of equation (2) we put:

$$\frac{(\hbar/2\pi)^2}{2m} = \left(\frac{(\hbar/2\pi)^3 |G|}{c}\right)^{1/2} \tag{3}$$

and obtains

$$m = \frac{1}{2} M_P$$

with new form of the equation (2)

$$i(\hbar/2\pi)\frac{\partial T}{\partial t} = \left(\frac{(\hbar/2\pi)^3 |G|}{c^5}\right)^{1/2} \frac{\partial^2 T}{\partial t^2} - \frac{(\hbar/2\pi)^2}{2m} \nabla^2 T. \tag{4}$$

Equation (4) is the quantum telegraph equation discussed in paper [4]. To clarify the physical nature of the solution of equation (4) we will discuss the diffusion approximation, *i.e.* we omit the second time derivative in equation (4) and obtain

$$i(\hbar/2\pi)\frac{\partial T}{\partial t} = -\frac{(\hbar/2\pi)^2}{2m} \nabla^2 T. \tag{5}$$

Equation (5) is the Schrödinger type equation for the temperature field in a universes with $G < 0$.

Both equation (5) and diffusion equation:

$$\frac{\partial T}{\partial t} = \frac{(h/2\pi)^2}{2m} \nabla^2 T \tag{6}$$

are parabolic and require the same boundary and initial conditions in order to be ``well posed''.

The diffusion equation (6) has the propagator [10]:

$$T_D(\vec{R}, \Theta) = \frac{1}{(4\pi D\Theta)^{3/2}} \exp\left[-\frac{R^2}{2\pi(h/2\pi)\Theta}\right], \tag{7}$$

where

$$\vec{R} = \vec{r} - \vec{r}', \quad \Theta = t - t'.$$

For equation (5) the propagator is:

$$T_s(\vec{R}, \Theta) = \left(\frac{M_P}{2\pi(h/2\pi)\Theta}\right)^{3/2} \exp\left[-\frac{3\pi i}{4}\right] \cdot \exp\left[\frac{iM_P R^2}{2\pi(h/2\pi)\Theta}\right] \tag{8}$$

with initial condition $T_s((R)\, 0) = \delta(R)$

In equation (8) $T_s((R), \Theta)$ is the complex function of R and $\Theta$. For anthropic observers only the real part of $T$ is detectable, so in our description of universe we put:

$$\mathrm{Im} T(\vec{R}, \Theta) = 0. \qquad (9)$$

The condition (9) can be written as (bearing in mind formula (8)):

$$\sin\left[-\frac{3\pi}{4} + \left(\frac{R}{L_P}\right)^2 \frac{1}{4\tilde{\Theta}}\right] = 0, \qquad (10)$$

where $L_P = \tau_P c$ and $\tilde{\Theta} = \Theta/\tau_P$. Formula (10) describes the discretization of $R$

$$R_N = [(4N\pi + 3\pi) L_P]^{1/2} (tc)^{1/2}, \qquad (11)$$

$$N = 0, 1, 2, 3\ldots$$

In fact from formula (11) the Hubble law can be derived

$$\frac{\dot{R}_N}{R_N} = H = \frac{1}{2t}, \quad \text{independent of } N. \qquad (12)$$

In the subsequent we will consider $R$ (11), as the space-time radius of the $N-$ universe with ``atomic unit'' of space $L_P$.

It is well known that idea of discrete structure of time can be applied to the ``flow'' of time. The idea that time has ``atomic'' structure or is not infinitely divisible, has only recently come to the fore as a daring and sophisticated hypothetical concomitant of recent investigations in the physics elementary particles and astrophysics. Descartes [12]. The shortest unit of time, atom of time is named *chronon* [13]. Modern speculations concerning the *chronon* have often be related to the idea of the smallest natural length is $L_P$. If this is divided by velocity of light it gives the Planck time $\tau_P=10^{-43}$ s, *i.e. the chronon* is equal $\tau_P$. In that case the time $t$ can be defined as

$$t = M\,\tau_P, \qquad M=0, 1, 2, ... \qquad (13)$$

Considering formulae (8) and (13) the space-time radius can be written as

$$R(M, N) = (\pi)^{1/2} M^{1/2} \left( N + \frac{3}{4} \right)^{1/2} L_P, \qquad M, N = 0, 1, 2, 3, \ldots \qquad (14)$$

Formula (14) describes the discrete structure of space-time. As the $R(M, N)$ is time dependent, we can calculate the velocity, $v = dR/dt$, *i.e.* the velocity of the expansion of space-time

$$v = \left(\frac{\pi}{4}\right)^{1/2} \left(\frac{N+3/4}{M}\right)^{1/2} c, \qquad (15)$$

where $c$ is the light velocity. We define the acceleration of the expansion of the space-time

$$a = \frac{dv}{dt} = -\frac{1}{2}\left(\frac{\pi}{4}\right)^{1/2} \frac{(N+3/4)^{1/2}}{M^{3/2}} \frac{c}{\tau_P}. \tag{16}$$

Considering formula (16) it is quite natural to define Planck acceleration:

$$A_P = \frac{c}{\tau_P} = \left(\frac{c^7}{(h/2\pi)G}\right)^{1/2} = 10^{51} \text{ ms}^{-2} \tag{17}$$

and formula (16) can be written as

$$a = -\frac{1}{2}\left(\frac{\pi}{4}\right)^{1/2} \frac{(N+3/4)^{1/2}}{M^{3/2}} \left(\frac{c^7}{(h/2\pi)G}\right)^{1/2}. \tag{18}$$

In table I the numerical values for $R$, $v$ and $a$ are presented. It is quite interesting that for $N, M \to \infty$ the expansion velocity $v < c$ in complete accord with relativistic description. Moreover for $N, M \gg 1$ the $v$ is relatively constant $v \sim 0.88\ c$. >From formulae (11) and (15) the Hubble parameter $H$, and the age of our Universe can be calculated

$$v = HR, \quad H = \frac{1}{2M\tau_P} = 5\cdot 10^{-18} \text{ s}^{-1},$$

$$T = 2M\tau_P = 2\cdot 10^{17} \text{ s} \sim 10^{10} \text{ years}, \tag{19}$$

which is in quite good agreement with recent measurement [15,16,17].

As is well known in de Sitter universe the cosmological constant $\Lambda$ is the function of R , radius of the Universe,

$$\Lambda = 3/R^2 \qquad (20)$$

Substituting formula ( 11) to formula (20) we obtain

$$\Lambda = 3/(\pi N^2 L_p^2), \quad N=0,1,2\ldots\ldots \qquad (21)$$

The result of the calculation of the radius of the Univers, *R*, the accelaration of the spacetime, *a*, and the cosmological constant, $\Lambda$ are presented in Figs. 1,2,3,4 for different values of number N. As can be easily seen the values of a and R are in very good agreement with observational data for prsent Epoch. As far as it is concerned cosmological constant $\Lambda$ for the firs time we obtain , the history of cosmological constant from thr Biginning to the present Epoch

# Figures 1,2,3,4

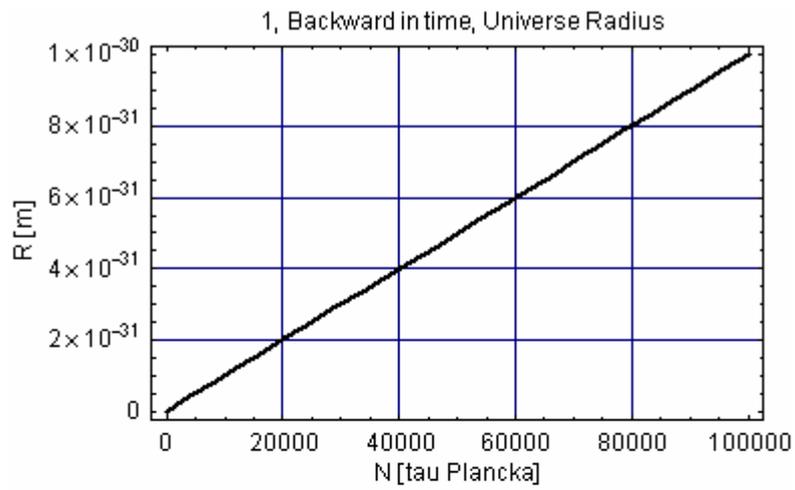

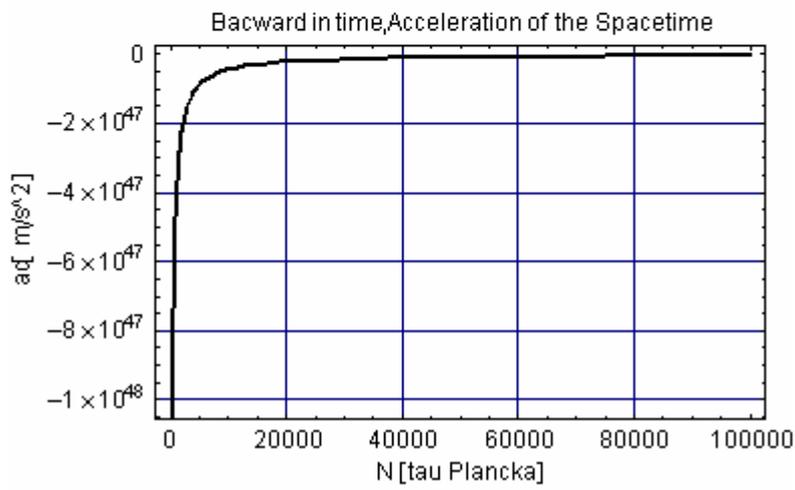

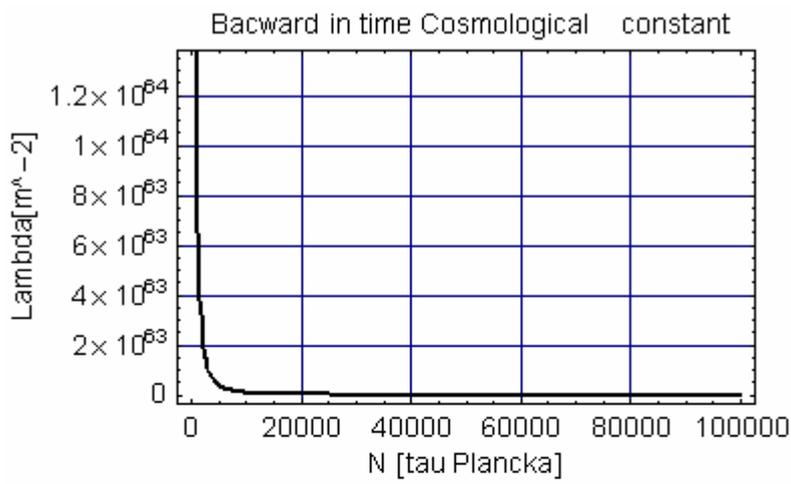

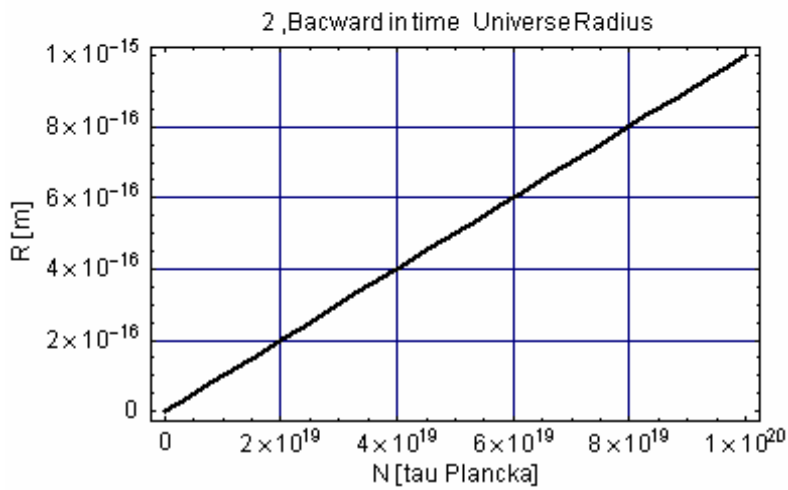

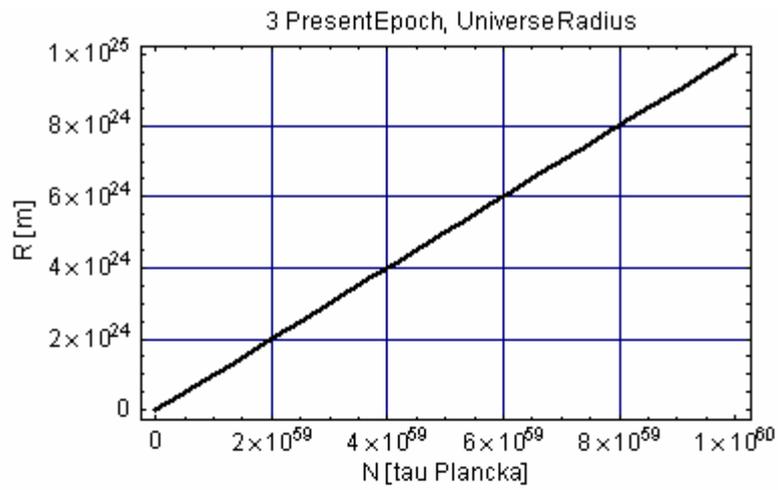

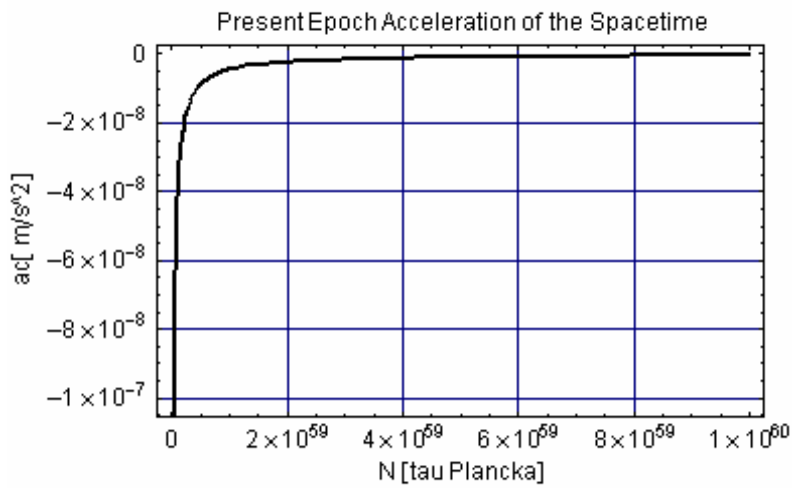

- Graphics -

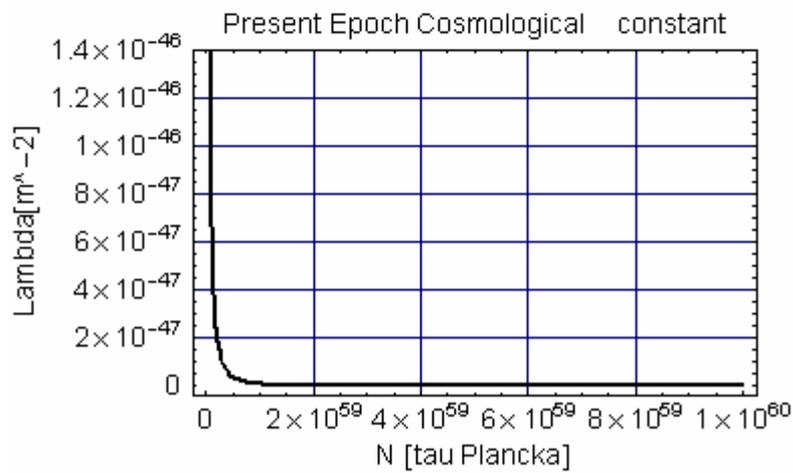

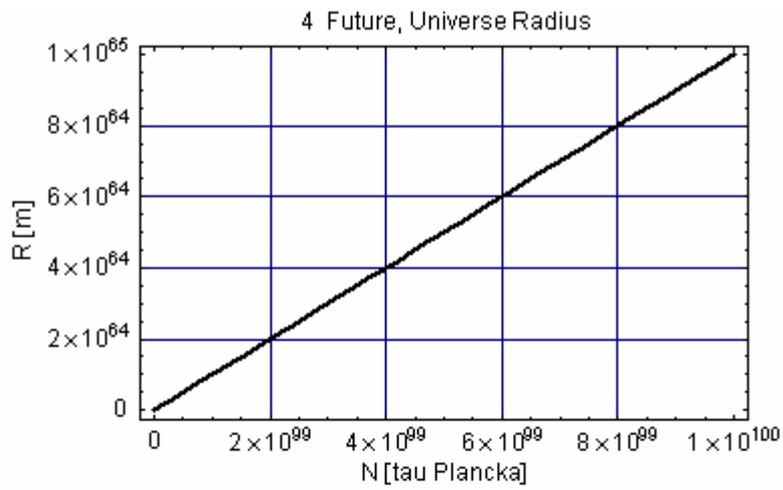

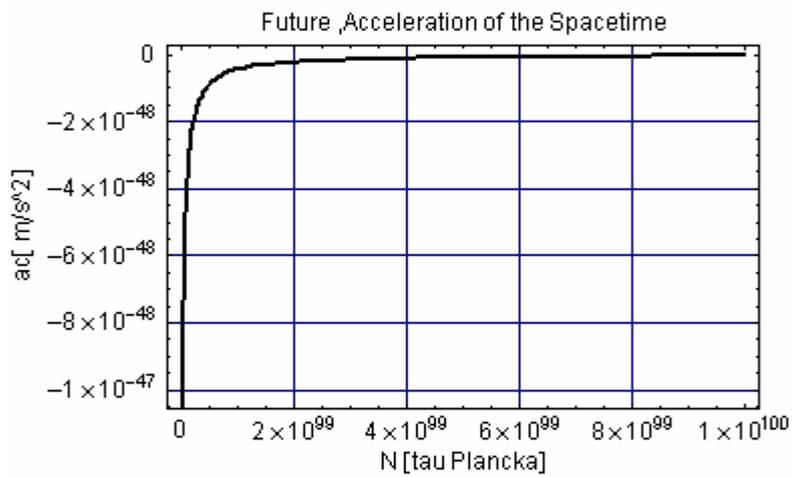

- Graphics -

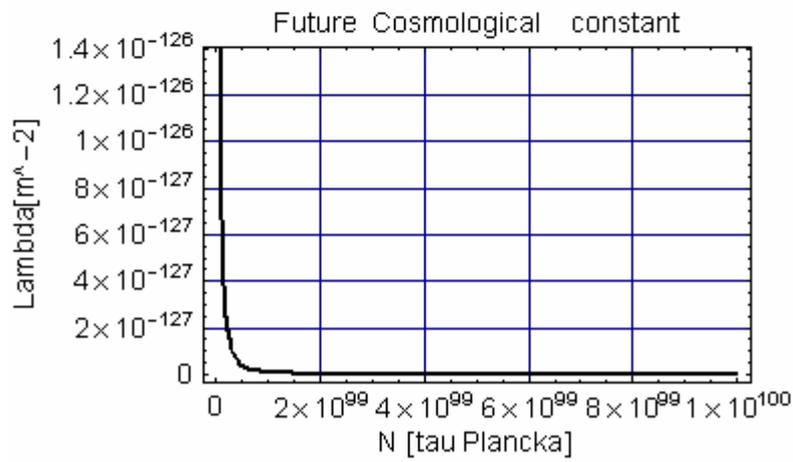

References

[1]